\documentclass[pra,twocolumn,superscriptaddress,floatfix,amsmath,amssymb]{revtex4-1}
\usepackage{amssymb}
\usepackage{epsfig}
\usepackage{amsmath}
\usepackage{amsfonts}
\usepackage{graphicx}
\usepackage{epsfig}
\graphicspath{{figure file fold path}}
\usepackage[dvipsnames,usenames]{color}
\usepackage[symbol*]{footmisc}

\usepackage{soul}

\begin{document}
\title{
Superthermal light emission and nontrivial photon statistics in small lasers}

\author{T. Wang}
\altaffiliation[Currently at: ]{Department of Electronic Engineering and Information Science, Hangzhou Dianzi University, Hangzhou, China.}

\author{
\hspace{-1.5mm}\normalfont\textsuperscript{, \ddag ~}
D. Aktas} 
\altaffiliation[Currently at: ]{Quantum Engineering Technology Labs, H. H.
Wills Physics Laboratory and Department of Electrical and Electronic
Engineering, University of Bristol, Bristol BS8 1FD, UK.\\
\hglue -3mm \normalfont\textsuperscript\ddag \ These Authors have equally contributed to this work.}

\author{
\hspace{-1.5mm}\normalfont\textsuperscript{, \ddag ~}
O. Alibart}
\author{\'E. Picholle}
\affiliation{Universit\'e C\^ote d'Azur, Institut de Physique de Nice (INPHYNI), CNRS UMR 7010, Nice, France}
\author{G.P. Puccioni}
\affiliation{Istituto dei Sistemi Complessi, CNR, Via Madonna del Piano 10,
I-50019 Sesto Fiorentino, Italy}
\author{S. Tanzilli}
\affiliation{Universit\'e C\^ote d'Azur, Institut de Physique de Nice (INPHYNI), CNRS UMR 7010, Nice, France}
\author{G.L. Lippi}
\affiliation{Universit\'e C\^ote d'Azur, Institut de Physique de Nice (INPHYNI), CNRS UMR 7010, Nice, France}
\email{gian-luca.lippi@inphyni.cnrs.fr}

\date{\today}

\begin{abstract}
Photon statistical measurements on a semiconductor microlaser, obtained using single-photon counting techniques, show that a newly discovered spontaneous pulsed emission regime possesses superthermal statistical properties.  The observed spike dynamics, typical of small-scale devices, is at the origin of an unexpected discordance between the probability density function and its representation in terms of the first moments, a discordance so far unnoticed in all devices.  The impact of this new dynamics is potentially large, since coincidence techniques are presently the sole capable of characterizing light emitted by nanolasers.
\end{abstract}

\maketitle 

\section{Introduction}
Photon statistics, a genuine workhorse of laser physics and quantum optics, now stands as a high-performance toolbox for the characterization of optical sources.  These range from single photon emitters~\cite{Somaschi2016} for entanglement-based secure communications~\cite{Aktas2016}, to coherent optical sources approaching the limit of a few (or single) electromagnetic cavity modes including micro- and nanoscale lasers.

Small-scale emitting devices are regarded as key components in future photonic integrated circuits~\cite{Smit2012} and in ever faster optical communications and data processing~\cite{Mayer2015,Yokoo2017,Takiguchi2017}. A thorough knowledge of the statistical properties of photons in the different environments is fundamental for a correct understanding of their features, especially when they strongly interact with a thermal bath (e.g., in semiconductor-based devices). For very small lasers, photon statistics is currently the sole technique capable of identifying the coherence threshold~\cite{Strauf2006,Takiguchi2015,Hayenga2016,Pan2016} and of characterizing the whole transition away from thermal emission~\cite{Takemura2012,Lebreton2013}.

The terms ``large" and ``small", often describing device size, are defined here in connection with the fraction of spontaneous emission coupled into the lasing mode, $\beta$.  Since at threshold the average photon number $\langle n_{th} \rangle \approx \beta^{-1/2}$~\cite{Rice1994}, assuming Poisson statistics the fluctuations (standard deviation) become $\sigma_{n_{th}} = \beta^{-1/4}$ and the relative fluctuations scale as $\frac{\sigma_{n_{th}}}{\langle n_{th} \rangle} = \beta^{1/4}$~\cite{Rice1994}.

By convention, we will consider lasers with a 10\% to 30\% noise contribution (spontaneous emission) -- thus with $10^{-4} \le \beta \le 10^{-2}$ -- as mesoscopic, and lasers with a lower noise (typically, 1\% for $\beta \approx 10^{-8}$) as macrolasers. Conversely, above 30\%, the noise contribution is very substantial, and the laser belongs to the nanoscale.  The limiting case, $\beta \rightarrow 1$, considered nowadays as the main goal for technological developments, is characterized by relative fluctuations which tend to 100\%.  Table~\ref{laserscale} provides a summary of this classification.

\begin{table} [h]
\begin{center}
\caption{Laser scale classification based on~\cite{Rice1994}.}  
\begin{tabular}{c c c}
Laser scale & $\beta$ range & Relative fluctuations range \\ \hline
Macrolaser & $0 \lessapprox \beta \lessapprox 10^{-4}$ & $0 \lessapprox \frac{\sigma_{n_{th}}}{\langle n_{th} \rangle} \lessapprox 0.1$ \\
Mesolaser & $10^{-4} \lessapprox \beta \lessapprox 10^{-2}$ & $ 0.1 \lessapprox \frac{\sigma_{n_{th}}}{\langle n_{th} \rangle} \lessapprox 0.3$ \\
Nanolaser & $10^{-2} \lessapprox \beta \lessapprox 1$ & $0.3 \lessapprox\frac{\sigma_{n_{th}}}{\langle n_{th} \rangle} \lessapprox 1$ \\\hline
\end{tabular}
\label{laserscale}
\end{center}
\end{table}

It is important to remark that semiconductor-based nanolasers intrinsically belong to the dynamical Class B~\cite{Tredicce1985} -- due to the slower temporal response of their material compared to the photon lifetime in the cavity -- while laser photon statistics has been developed in the 1960s and 1970s for Class A devices (in general atomic lasers) where the slowest variable is the electromagnetic field (cf., e.g.,~\cite{Arecchi1971}; also \cite{Takemura2019} for a modern examination of class A nanodevices).  Originally developed for macroscopic lasers~\cite{Paoli1988,Ogawa1990}, Class B laser photon statistics, which takes into account the non-markovian memory effects arising from the slow material response, was experimentally tested at the microscale on a solid-state microcavity~\cite{Lien2001}, taken as a substitute for a semiconductor microcavity~\cite{Woerdman2001} whose typical time constants presented, at the time, impossible technical challenges. The extreme Class B nature of the solid-state device~\cite{Lien2001} -- identified by $\Lambda \beta > 1$, where $\Lambda = \frac{\Gamma_c}{\gamma_{\parallel}}$ ($\Gamma_c$ cavity losses, $\gamma_{\parallel}$ carrier relaxation rate), allowed the authors to extend the validity of the Class B photon statistics to all laser sizes~\cite{Lien2001,vanDruten2000}.

The purpose of this paper is to experimentally verify this long-standing conclusion on a Vertical Cavity Surface Emitting Laser (VCSEL) with a moderate fraction of spontaneous emission coupled into the lasing mode, $\beta \approx 10^{-4}$, and $\Lambda \approx 10^{2}$. By investigating both the Probability Density Function (histogram P (n) for the photon number n obtained from linear measurements) and the autocorrelation function $g^{(2)}(0)$, we are able to shine a new light on the photon statistics of Class B lasers. 

In the following, we will show the existence of an emission regime with superthermal correlations accompanying the intrinsic photon emission dynamics of a {\it single mode} micro-cavity laser in the transition from incoherent to coherent emission. 
While superthermal statistics, recently observed in nanoscale two-mode systems~\cite{Schlottman2017,Marconi2018}, is known to appear in concomitance with mode-switching~\cite{Redlich2016}, single-mode emission is expected to monotonically evolve from thermal to Poisson statistics as the field acquires coherence.  The new regime we observe, inconsistent with the predicted probability density function (PDF) for a Class B laser~\cite{Paoli1988}, stems from the intrinsic threshold dynamics of the microdevice, which is so far unaccounted for in analytical models of laser emission.

\section{Experimental setup}

\begin{figure}[htbp]
\centering
\includegraphics[width=\linewidth,clip=true]{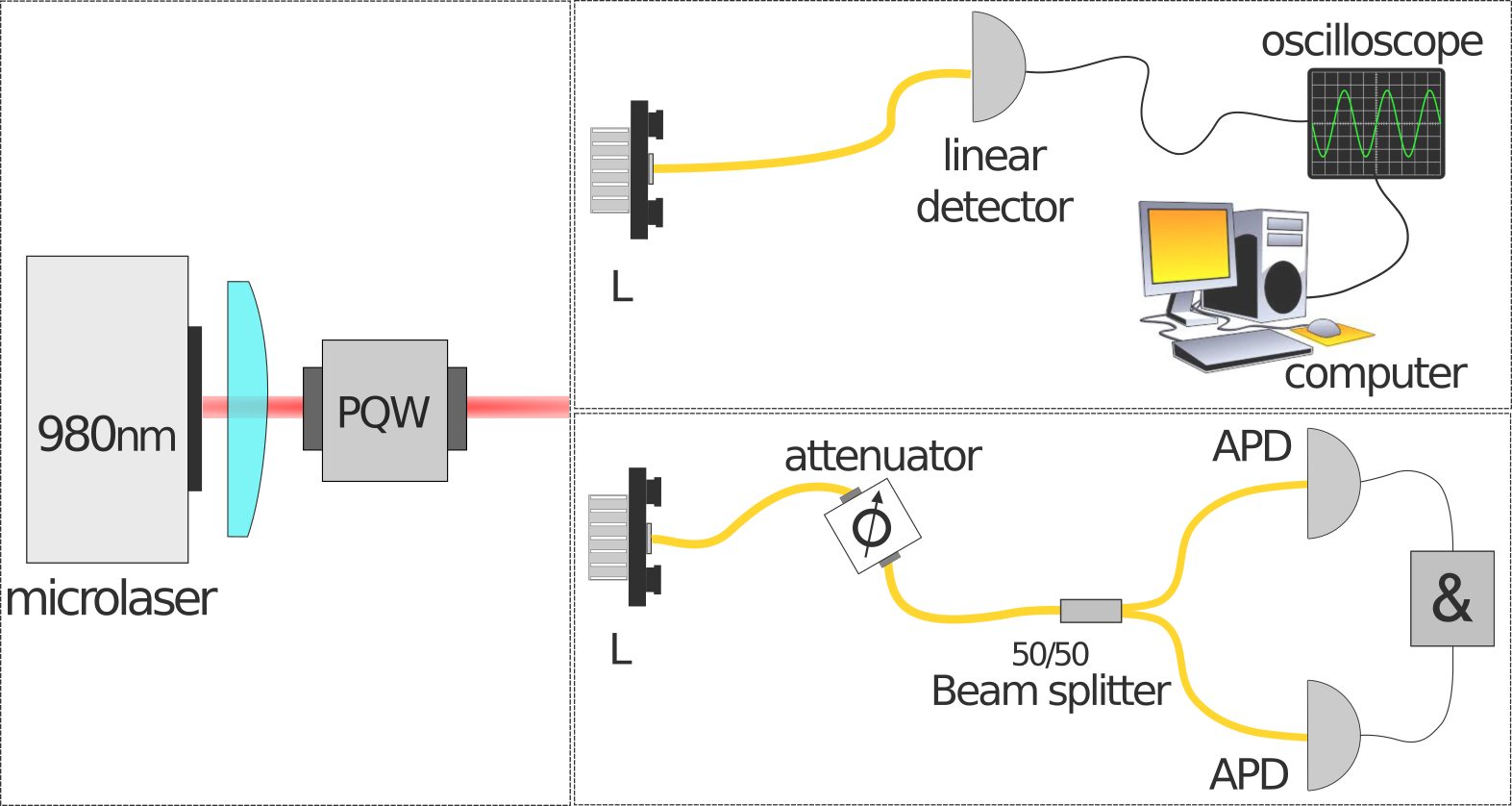}
\caption{
Experimental setup. The laser (Thorlabs 980 VCSEL -- $\lambda = 980$ nm --, threshold current $i_{th} \approx 1$ mA) is temperature-stabilized and supplied by a low-noise DC source (Thorlabs LDC200VCSEL). \textbf{Top:} The collimated laser output traverses an optical isolation stage (PQW), is collected by lens L and couples through a fiber into a 10 GHz photodetector (Thorlabs PDA8GS). The signal, digitized by a LeCroy Wavemaster 8600 oscilloscope ($6$ GHz analog bandwidth, $10^6$ points/trace), is stored for treatment. \textbf{Bottom:} The photon counting apparatus consists of an attenuator (Thorlabs VOA980 FC), a 50/50 fiber beamsplitter and 2 single photon detectors (APD idQuantique id100) with $\sim$ 40\,ps jitter. The AND gate (\&) is made with a TAC (Ortec 567) with $\sim$ 15\,ps timing resolution. 
}
\label{setup1}
\end{figure}

As depicted in Fig.~\ref{setup1}, the experiment consists of:  (1) the measurement of $P(n)$, performed with a linear detector; and (2) the direct measurement of $g^{(2)}(0)$ by single-photon interferometry using a standard Hanbury-Brown and Twiss (HBT) setup.

\begin{figure}[htbp]
\centering
\includegraphics[width=\linewidth,clip=true]{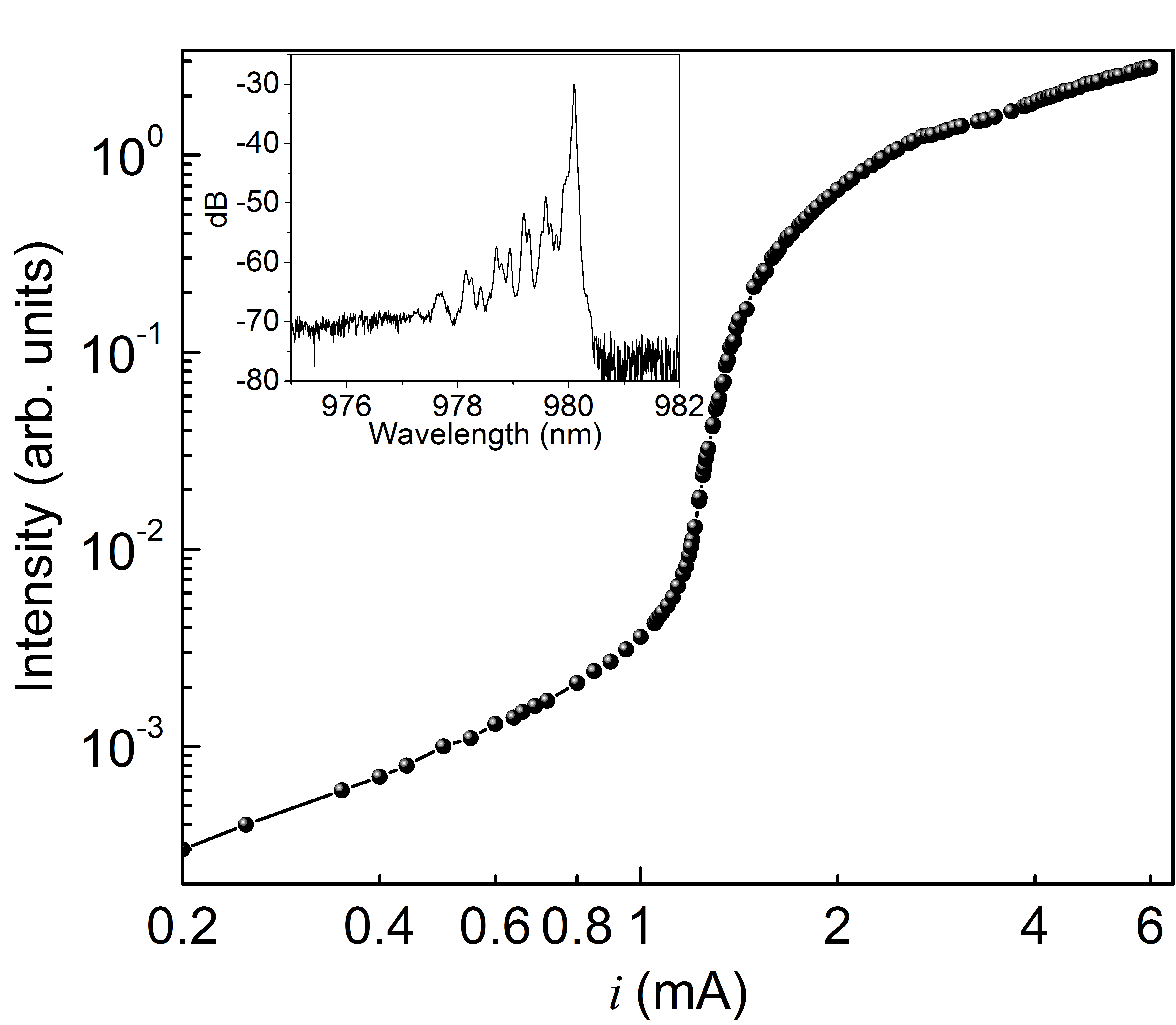}
\caption{
Laser input-output characteristics in double-logarithmic scale as a function of pumping current.  The inset shows a typical optical spectrum as it appears at $i \approx 1.25 mA$.
}
\label{IVcurve}
\end{figure}

The laser source is illustrated on the left panel of Fig.~\ref{setup1}. The choice of a VCSEL microcavity results from careful considerations and is crucial for the success of the investigation. It is a commercial VCSEL (Thorlabs VCSEL-980), emitting at the nominal wavelength $\lambda = 980$ nm and based on GaAs semiconducting junctions.  The maximum pump current allowed for this laser is $i_{max} = 10 mA$ while the maximum output power that can be obtained is $P_{max} \approx 1.85 mW$.  Two distinct devices of this series have been used for the experiment, obtaining consistent results. The laser temperature is set at $T_L = 25.0^{\circ}C$.  The active temperature stabilization comes from a home-built apparatus, capable of better than $0.1^{\circ}C$.  In the current range investigated, the laser emits along a single linear polarization ($> 20$dB suppression ratio), devoid of switching.  The typical input-ouput curve, in double logarithmic scale, consistent with a $\beta \approx 10^{-4}$~\cite{Wang2015} is shown in Fig.~\ref{IVcurve}.  Its inset shows a representative optical emission spectrum in the lower part of the steep region of the laser response:  the strong component is relatively narrow-band, while other emission peaks exist with contributions at least 20dB below the maximum (at $\lambda \approx 980$ nm).  This spectral characteristics is quite standard for these types of devices and evolves towards a narrower peak (with additional relative suppression of other wavelenghts) as lasing sets in.

Before entering the specific detection system, the signal output by the laser traverses an {\it optical isolation stage} (PQW:  Polarization beamsplitter-Quarter-Wave-plate ensemble) to strongly suppress backreflections into the device which may perturb its dynamical operation.

The top-right panel of Fig.~\ref{setup1} illustrates the classical part of the experimental setup.
The signal coming from the laser is coupled through a lens L into a multimode optical fiber and reaches a fast, amplified photodetector (Thorlabs PDA8GS) with bandwidth $9.5 GHz$.  The electrical signal is digitized by a LeCroy Wave Master 8600A oscilloscope with $6 GHz$ analog bandwidth and up to $5 \times 10^6$ sampled points per trace. The detector is supplied by a dc 12V battery, rather than by the power supply provided by the Manufacturer, to reduce the amount of electrical noise coming from the lines.

The bottom-right panel of Fig.~\ref{setup1} illustrates the quantum part of the experimental setup.
The signal coming from the laser is coupled through a lens L into a single-mode fiber going through an optical power attenuator (Thorlabs VOA980-FC) thus allowing to vary the attenuation up to $50 \rm \, dB$ in order to ensure that the input power going to the detecors is not too high. The signal is then split in two, thanks to a fiber beamsplitter (Thorlabs FC980-50B-FC), and each output port is connected to a single photon detector (APD IdQuantique id100) with $40 \rm \,ps$ jitter for precise coincidence measurements.
The electronics use to register coincidences is a time-to-amplitude converter and a single channel analyser (TAC/SCA ORTEC model 567). It measures the time interval between a start and stop input pulse, generates an analog output pulse proportional to the measured time, and provides built-in single channel analysis of the analog signal. The jitter introduced by this electronics depends on the settings and with a coincidence window set at $100 \rm \,ns$ this jitter is of the order of $15 \rm \, ps$.  In both detection schemes the total signal is recorded, without any polarization selection.  

\section{Techniques}

This investigation rests on the two detection schemes illustrated above and on the use of a fully stochastic numerical procedure to obtain predictions which are on the one hand necessary to compare the data to theoretical predictions, on the other hand to interpret the results.  The strong relative fluctuations typical of small lasers~\cite{Rice1994} render the standard modeling based on rate Equations~\cite{Coldren2012}, augmented by Langevin noise, invalid~\cite{Lippi2019}. 

Sections~\ref{quantum} and \ref{SS} present the technical aspects of the quantum coincidence measurements and illustrate the principle of the Stochastic Simulator, respectively.  The measurements obtained from the linear detector are straightforward and have been amply discussed elsewhere~\cite{Wang2015}.

\subsection{Quantum coincidence techniques}\label{quantum}

One typically determines $g^{(2)}(\tau)$ of a beam of light using a beamsplitter and measuring the correlation between the reflected and transmitted output intensities. In an experiment, one does not directly measure the intensity, hence the need to relate the expression for $g^{(2)}(0)$ to experimentally measured quantities. It can be shown that when $g^{(2)}(0)$ is measured with photodiodes, it is written in terms of probabilities of photodetections as $g^{(2)}(0) = P_{tr}/P_tP_r$, where $P_t$ (resp. $P_r$) is the probability of a photodetection at detector $T$ (resp. $R$) in a short interval $\Delta t$ and $P_{tr}$ is the joint probability of detection of an event occurring at both detectors in the same interval.

The autocorrelation function is obtained through coincidence measurements using two single photon detectors (Fig.~\ref{setup1}, bottom), recording with a time-to-amplitude-converter the time distribution of coincidences between the two detectors for different values of the injection current.

The single channel analyser provide us with an histogram of coincidences which is directly related to the autocorrelation function $g^{(2)}(\tau)$. 
We assume that, in a given integration time $\Delta T$, the detector registers $n_t$ photons in the beamsplitter transmission and $n_r$ photons in reflection. We therefore calculate the second order correlation function,
\begin{eqnarray}
g^{(2)}(\tau)= \frac{M_c}{n_t n_r} \cdot (\frac{\Delta T}{\Delta t}) \, ,
\end{eqnarray}
where $M_c$ is the number of coincidence detection per time bin $\Delta t$, i.e., the resolution of the system used to measure coincidences.

\begin{figure}[h!]
\centering
\includegraphics[width=\linewidth,clip=true]{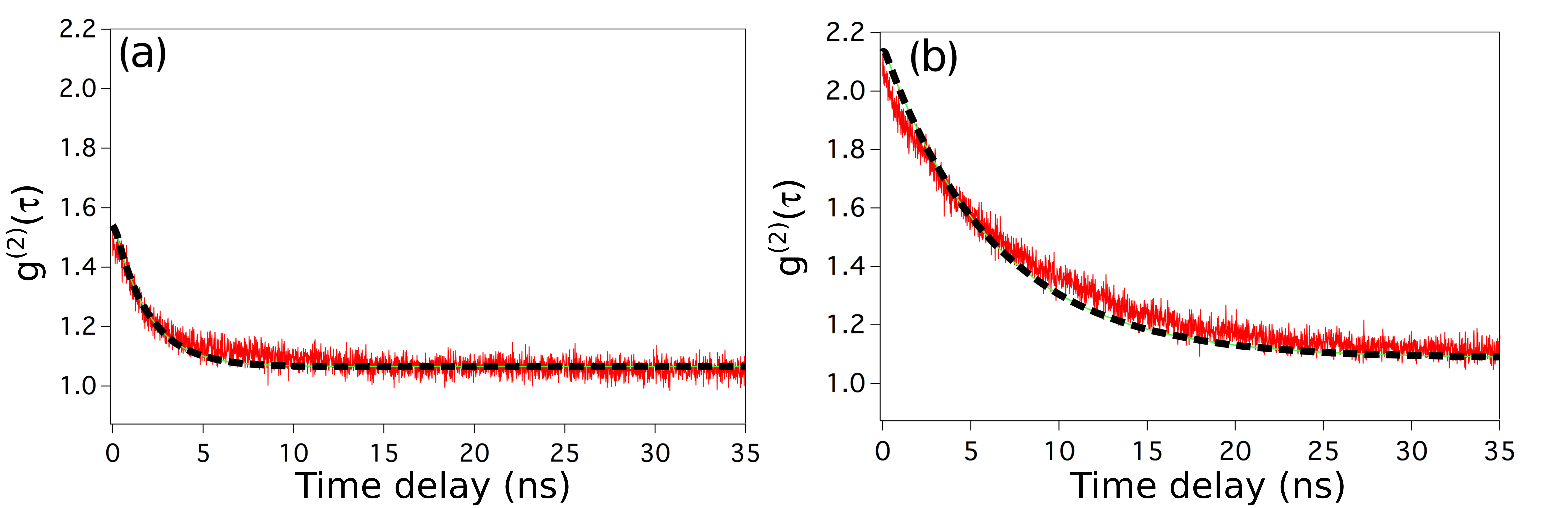}
\caption{
Autocorrelation function (solid line) for different values of the injection current. (a) $i=1.20 \rm \, mA$.  The coincidence counts (in red) converge to 1 outside the plotted window and allow us to read $g^{(2)}(0)\approx 1.5 $ with the help of the exponential decay fitting function (dashed line). (b) Similarly, for $i=1.30 \rm \, mA$ we obtain $g^{(2)}(0)\approx 2.1 $. 
}
\label{det-rawCoinc}
\end{figure}

We have collected histograms for several different values of the injection current:  examples for $i=1.20 \rm \,mA$ and $i=1.30 \rm \,mA$ are shown in Fig~\ref{det-rawCoinc}. Each data set is taken with different integration times ($\sim 10-20 \rm \, min$) since it is experimentally necessary to adapt its value, as the injection current is increased, in order to efficiently capture the dynamics without saturating the detectors. 

\subsection{Stochastic Simulator}\label{SS}

The numerical simulations are carried out with a novel way of emulating the laser dynamics through the basic physical processes on which the emission rests:  a semiclassical description of spontaneous and stimulated emission based on Einstein coefficients~\cite{Einstein1917} treated as fully stochastic processes.  The scheme, discussed in detail in~\cite{Puccioni2015} rests on a stochastic excitation of the upper laser level, on the evolution of carriers and of the different photon fields (stimulated, spontaneous on-axis and spontaneous off-axis), defined on the basis of recurrence relations and on the composition of the different events (pump processes, stimulated and spontaneous emission, transmission through the cavity output coupler) all treated as stochastic processes.  The consequence of this choice is that instead of treating average quantities, as in the usual differential approach, all physical variables have an instantaneous, intrinsically granular dynamics since all quantities are integer numbers, as in real lasers.  In addition to providing a short-term picture of the stochastic evolution of the laser and a good match to experimentally measured features (and dynamics), this model is capable of reproducing very closely the average quantities which are typical of the differential description.  The latter, instead, cannot be used to predict the dynamical behaviour at small scales due to the intrinsic violation of the necessary conditions for a Langevin representation of noise~\cite{Lippi2019}.

The probabilistic description of each event implies the intrinsic presence of noise in this model, since the randomness of the realization of each process introduces fluctuations without the need for the additional hypotheses necessary, for instance, to add Langevin noise terms to the continuous differential description (rate equations~\cite{Coldren2012}).  The automatic accounting for all kinds of spontaneous emission (on- and off-axis) also adds an element which render the model's predictions more realistic since they account for fluctuations in small devices, as realized very early on~\cite{Woerdman2001} and clearly emphasized by numerical simulations~\cite{Lebreton2013b}.  The match between the predictions of the Stochastic Simulator and experiments, in addition to reproducing the results coming from standard rate equations, reinforces the validity of this approach, particularly for the nontrivial features of the autocorrelation function~\cite{Wang2015}, which continuous models cannot predict.  Experimentally observed dynamical microlaser behaviour in the threshold region is also correctly predicted by this modeling approach~\cite{Wang2018,Wang2019}. 

This simulator has been used to obtain the numerical PDFs of Section~\ref{linear} (Fig.~\ref{compH-en}) and has shown the appearence of the peak in the autocorrelation discussed below (Section~\ref{quantummeas}, Fig.~\ref{g2}).  In addition, it offers the possibility of computing the dynamics for different values of the $\beta$-parameter and compare the trend to be expected as a function of effective cavity volume.  This feature will be later exploited to compare results obtained from devices of different sizes (cf. Section~\ref{dynprog}).

\section{Results}

The two different measurement techniques are used to explore two indicators:  the PDF, using the linear detection technique, and the second-order zero-delay autocorrelation, $g^{(2)}(0)$, using quantum coincidences.  Both indicators offer evidence of so-far unknown, or at least unidentified, aspects of statistical photon properties of light emitted by small cavity volume lasers.

\subsection{Fast linear detector measurement}\label{linear}

Manufacturer's specifications for the fast linear detector (InGaAs PIN detector, model PDA8GS), used to reconstruct the PDF, give a spectral responsivity $S(\lambda = 980 \rm nm) \approx 0.75A/W$.  According to manufacturer's specifications, coupling the built-in preamplifier into $50 \Omega$ gives an output signal $V_{out} = 460 \cdot P_{opt} \cdot S(\lambda)$, where 460 is the transimpedance amplification coefficient.  An electrical signal with measured amplitude $1 mV$ corresponds therefore to $P_{opt} \approx 3 \mu W$, which corresponds to the smallest signal which can be detected at the fastest timescale.  Conversion into photon number at $\lambda = 980 \rm nm$ can be achieved by computing the photon flux $\Phi_n \approx 1.5 \times 10^{13} s^{-1}$, i.e. approximately 1500 photons collected on the detector in the fastest interval it can resolve ($0.1 \rm ns$).  In the following, we present the experimental results in the measurement units, given that all conversions are affected by an additional uncertainty.

\begin{figure}[htbp]
\centering
\includegraphics[width=0.49\linewidth,clip,trim=0 50 40 0mm]{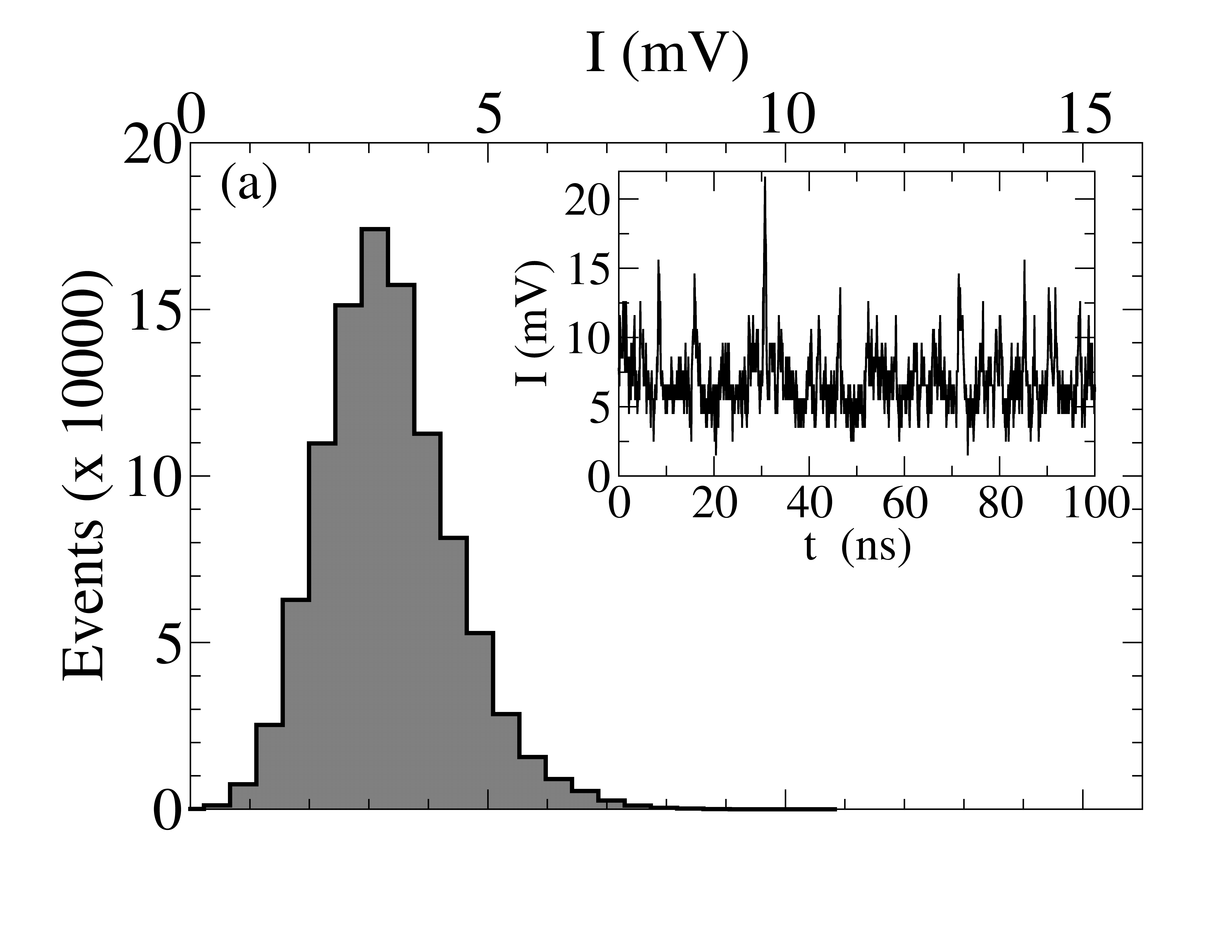}
\includegraphics[width=0.49\linewidth,clip,trim=20 50 0 0mm]{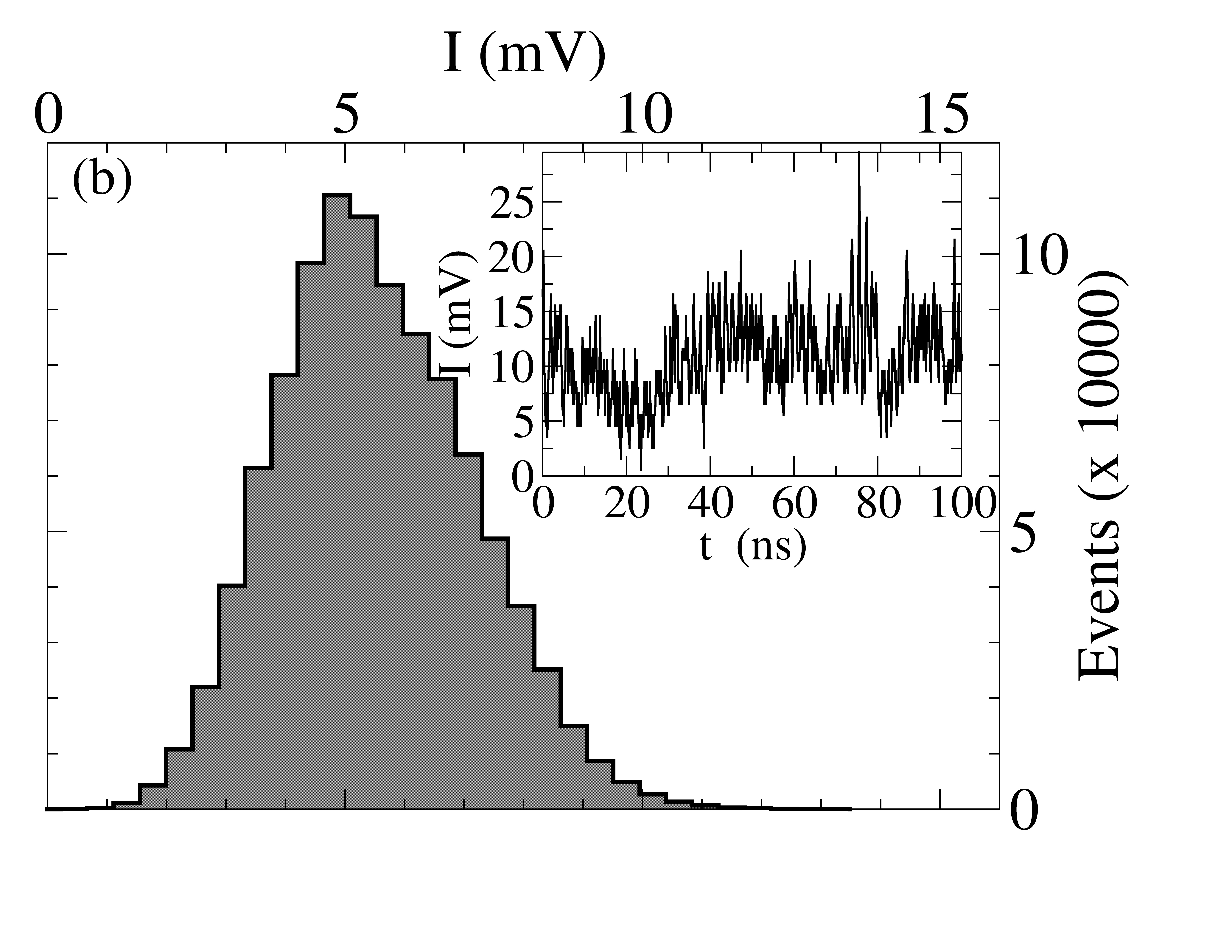}\\
\includegraphics[width=0.49\linewidth,clip,trim=0 0 40 25mm]{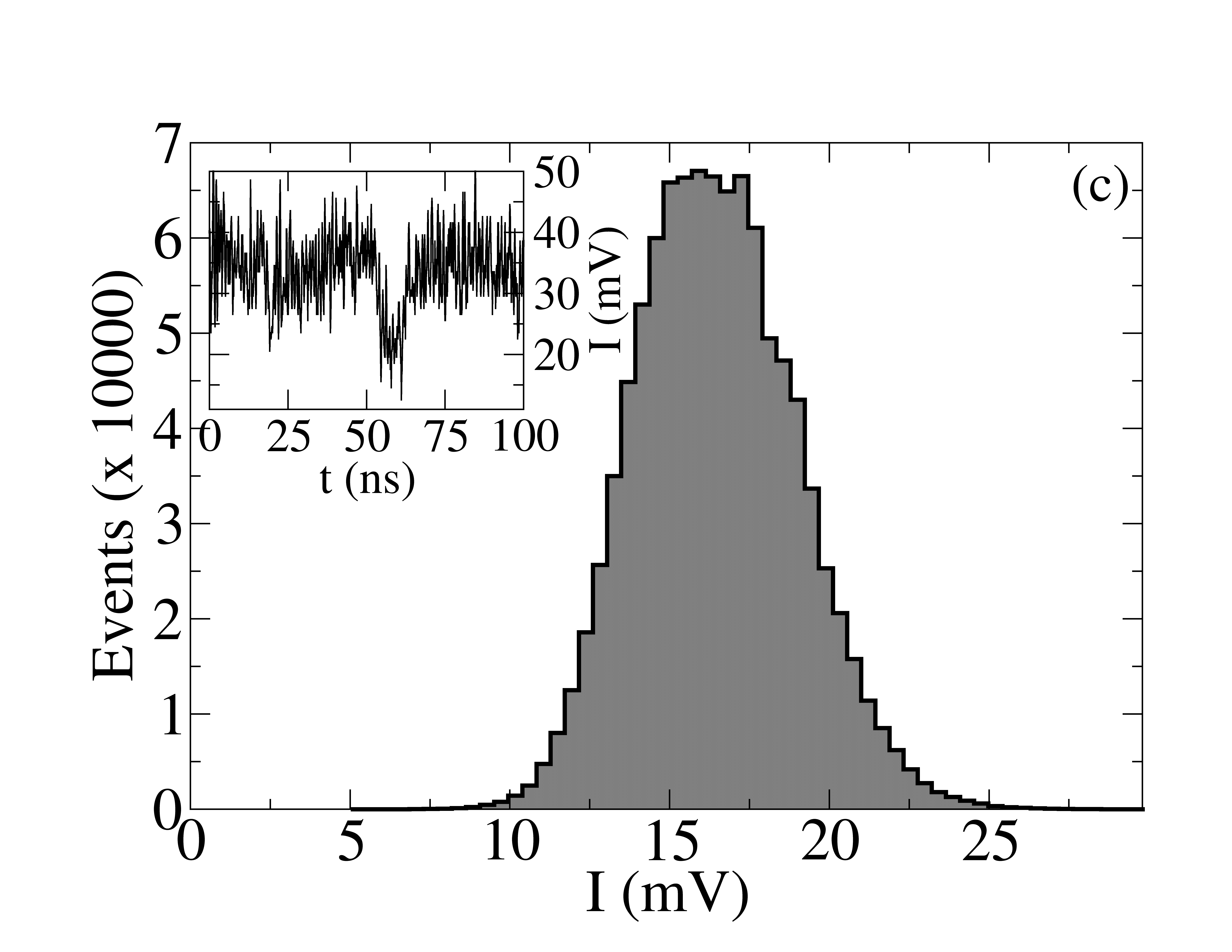}
\includegraphics[width=0.49\linewidth,clip,trim=20 0 0 25mm]{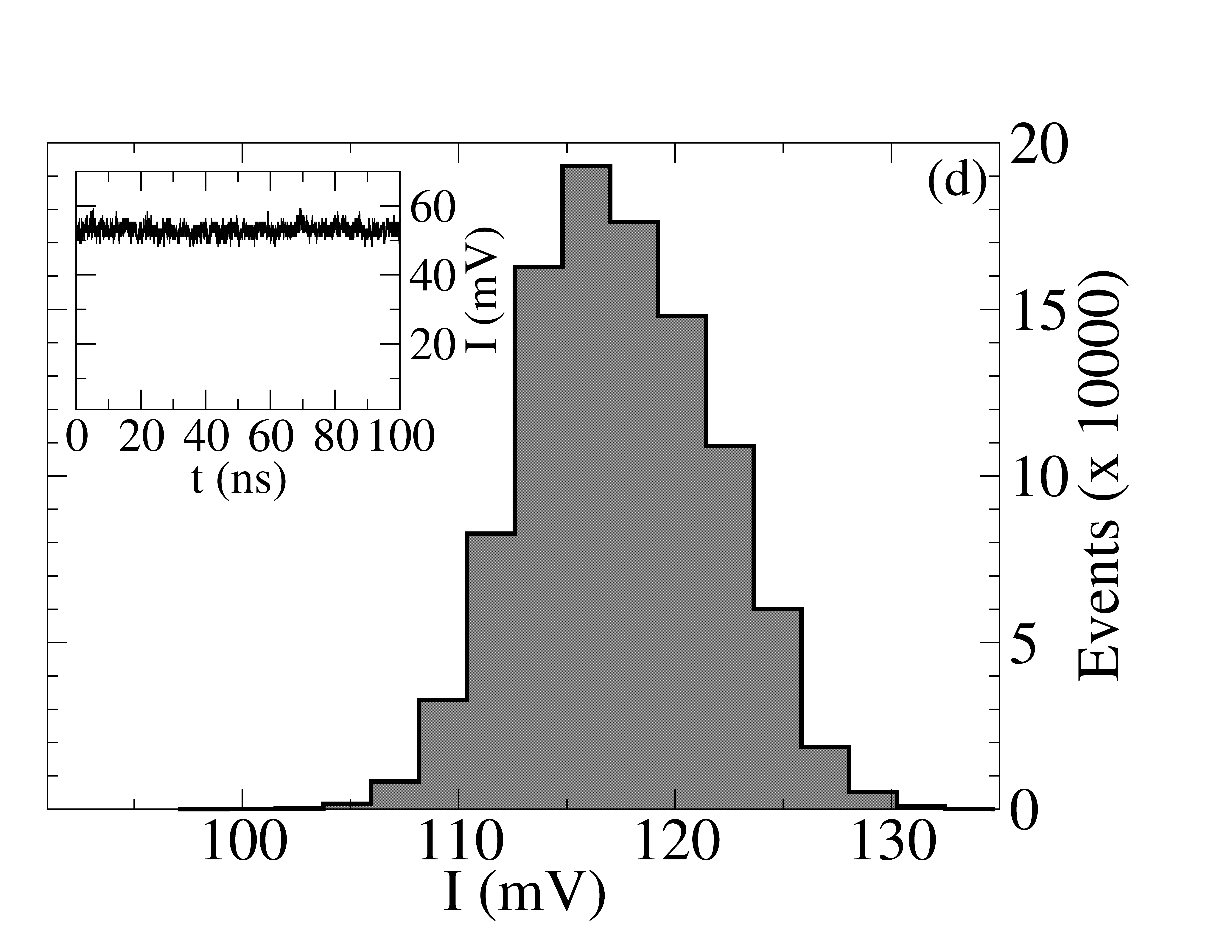}
\caption{
Experimental PDF on $10^6$ points collected with the linear setup (Fig.~\ref{setup1}, top).  $I$ represents the detected intensity (insets:  temporal traces).  Laser bias current:  (a) $i = 1.26$ mA (beginning of laser emission); (b) $i = 1.30$ mA (laser pulses); (c) $i = 1.45$ mA (fully developed intensity oscillations); (d) $i = 3.0$ mA (stable emission).  The thick PDF outline highlights the long tails of the distribution.
}
\label{expthisto}
\end{figure}

Fig.~\ref{expthisto} shows the histograms of the intensity distribution measured at significant values of pumping current (insets:  temporal traces).  The distributions show a distinct difference from Class A laser photon statistics, characterized by a large probability of recording no photons, and possess a signature similar to that of Class B devices~\cite{Lien2001}, whose distributions start from the origin even at the lowest pump values (as in Fig.~\ref{expthisto}a).

In addition, Fig.~\ref{expthisto} provides interesting information on the shapes of the distributions.  Panel (a), obtained closest to threshold, shows a narrow, peaked distribution with a tail towards large photon numbers (typical of a Class B laser PDF~\cite{Paoli1988}).  A slight increase in the laser pump current quickly widens the distribution, still maintaining an asymmetric tail to the right (Fig.~\ref{expthisto}b).  However, further increasing the pump current leads to a very noisy, but mostly continuous wave output (Fig.~\ref{expthisto}c), providing a much broader, nearly symmetric-looking distribution analysed in the following.  Finally, sufficiently far above threshold, we find the standard, low noise ($\lesssim 10\%$) above-threshold laser emission (Fig.~\ref{expthisto}d).

\begin{figure}[htbp]
\centering
\includegraphics[width=\linewidth,clip=true]{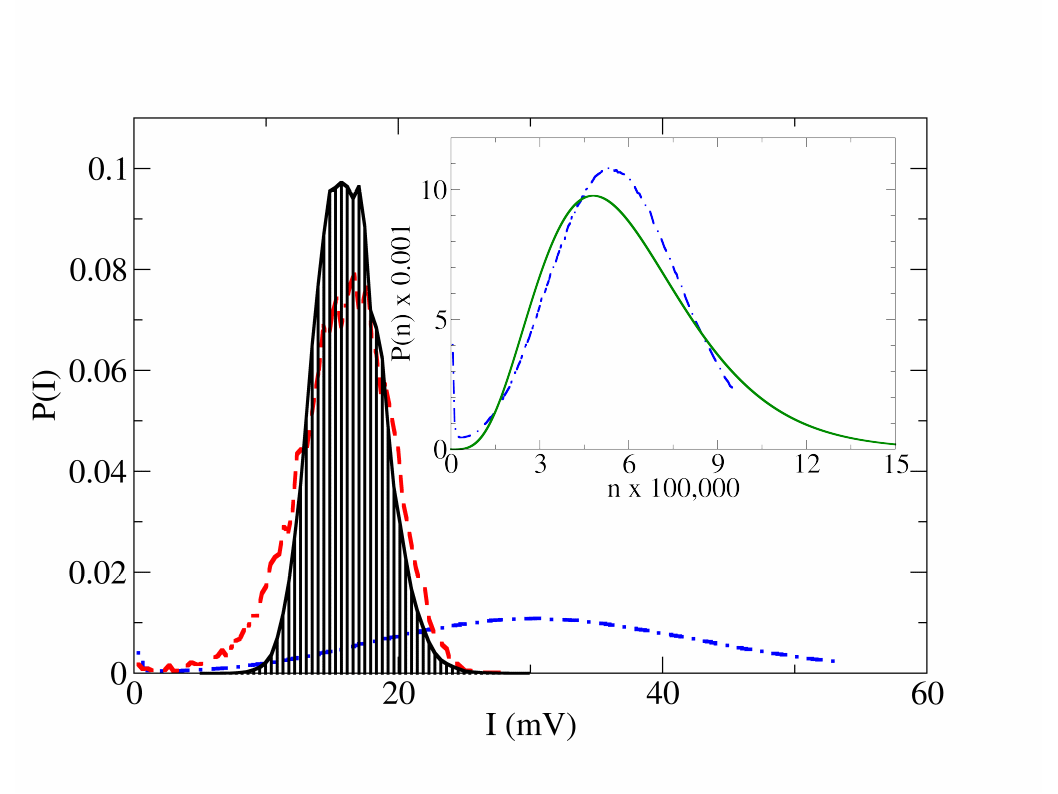}
\caption{Comparison between experimental distribution (hashed histogram, from Fig.~\ref{expthisto}c) and numerical simulations (SS) to highlight the detector's filtering action.  The dashed (red online) and the dot-dashed (blue online) lines represent the filtered and unfiltered numerical PDFs, respectively.  Inset:  Comparison between the theoretical $P(n)$ (solid line -- green online) obtained from~\cite{Paoli1988} and its numerical (SS) intracavity counterpart (dot-dashed line -- blue online).  The accumulation of events in the ``0" channel in the simulation is due to the absence of external noise, which results in the appearence of strictly zero photons after an occasional very large fluctuation.
}
\label{compH-en}
\end{figure}

The measured statistical distributions (Fig.~\ref{expthisto}) are strongly affected by the data acquisition chain bandwidth $f_B$ ($2 \pi f_B < \frac{\Gamma_c}{10}$).  Since it is impossible to deconvolve its effect on the measured statistical distribution, we resort to comparing our results to the photon number distribution numerically predicted by a Stochastic Simulator (SS)~\cite{Puccioni2015}, known to provide reliable predictions of the laser's dynamics~\cite{Wang2015}.  Comparison of the predicted statistics with and without the detection's filtering action offers a pertinent tool for bridging the gap between experimental observations and theoretical predictions~\cite{Paoli1988}.  This test is performed on the distribution of Fig.~\ref{expthisto}c, which corresponds to the very noisy laser output.  The predictions of the SS are extracted from a large data sample ($10^7$ points), with time binning $t_b$ ($0.1 \rm ns$) compared to the photon lifetime ($t_b  \ll \Gamma_c^{-1}$) to obtain the probability distribution represented by the dash-dotted line in Fig.~\ref{compH-en}.  We then mimic the action of the detector by binning the data on a time $\tilde{t}_b \approx f_B^{-1}$ (dashed curve).  The influence of the acquisition chain on $P(n)$ is dramatic:  the distribution is strongly narrowed, and thus raised, with its maximum at lower average photon numbers.  The figures display the statistics of output photons -- as in the experiment -- rather than the usually predicted intracavity photon distribution.  Comparison to the experimental distribution (filled histogram in Fig.~\ref{compH-en}) shows a good qualitative agreement between predicted and measured $P(n)$.  The numerical curves have been rescaled to account for the detector's sensitivity and the horizontal scale is given in mV to preserve the experimental units. 

Matching experimental and theoretical probability distributions requires an additional step.  The inset in Fig.~\ref{compH-en} reproduces the computed $P(n)$ (same dot-dashed curve as in the main figure) impinging on an ideal, non-bandwidth-limited detector, plotted together with the theoretical intracavity photon distribution (solid line, green online)~\cite{Paoli1988}.  The values for the two parameters characterizing the probability distribution are chosen to be $\overline{n} = 4 \times 10^5$ (for the probability distribution defined in~\cite{Lien2001}) consistent with both rate-equation models and the corresponding average value from the Stochastic Simulator, and $g^{(2)}(0) = 1.2$, inferred from  the zero-delay second-order autocorrelation measurements~\cite{Wang2015}, where we have taken into account the correction due to the detection bandwidth which reduces $g^{(2)}(0)$ by approximately one order of magnitude ($g^{(2)}(0)_{meas} - 1 \approx \frac{f_B}{\Gamma_c} \left[ g^{(2)}(0)_{true} - 1 \right]$, where $g^{(2)}(0)_{meas}$ stands for the measured value of the autocorrelation and $g^{(2)}(0)_{true}$ for its unfiltered counterpart~\cite{BoitierFabre2009}). 

The numerical histogram is sharply truncated (inset of Fig.~\ref{compH-en}).  This cutoff comes from the limited number of emitters and from the small amount of fluctuations in the carrier density tolerated around threshold:  $\sim 1\%$ is sufficient to turn off the laser.  In a solid-state microcavity, the photon number fluctuations supported by the laser are much larger due to low output coupler reflectivity ($R = 0.8$)~\cite{vanDruten2000}, resulting in much longer tails in the experimental PDF~\cite{Lien2001}. 

Overall, considering the significant error margin in the estimation of some parameters, such as cavity losses and detecton efficiency, the match is quite satisfactory.  We thus conclude, by comparing the curves in the main part and inset of Fig.~\ref{compH-en}, that the experimental photon distribution is compatible with the predictions of theoretical models for Class B lasers~\cite{Paoli1988}, and is consistent with the investigations of Ref.~\cite{Lien2001}. 

However, the current experimental results (Fig.~\ref{expthisto}), enabled by modern instrumentation, do not correspond to a standard (i.e. Class A) photon statistics, despite satisfying the criterion for a {\it weakly class B} device, namely $\beta \Lambda < 1$.  Thus, they invalidate the conclusion of~\cite{Lien2001}. This first unexpected results sheds new light onto the photon statistics of semiconductor microcavities.

\subsection{Quantum coincidence measurements}\label{quantummeas}

A more spectacular discrepancy appears when looking at the correlations measured with a Hanbury-Brown \& Twiss (HBT) setup in the single photon counting regime, whose sensitivity can reach the femtowatt level.  
The combination of strongly accrued sensitivity and larger measurement bandwidth enhances the fidelity in the reproduction of the laser's behaviour and, through the direct comparison between $P(n)$ and $g^{(2)}(0)$, sheds additional light on what is to be expected in nanolasers.

\begin{figure}[htbp]
\centering
\includegraphics[width=1\linewidth,clip=true]{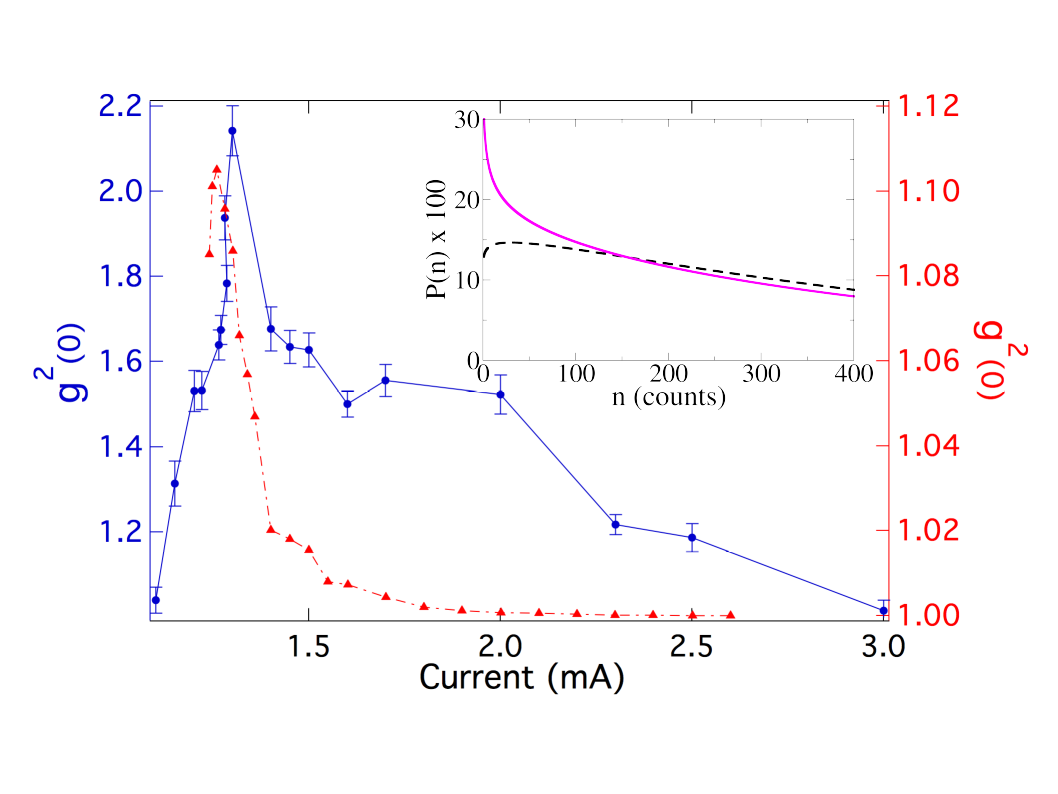}
\caption{
Scaled second-order autocorrelation ($g^{(2)}(\tau = 0)$) obtained with the HBT setup as a function of the injection current (dots -- blue online -- left vertical scale). For comparison, the previously reported measurements of $g^{(2)}(\tau = 0)$ ~\cite{Wang2015} obtained from linear detection (triangles -- red online -- right vertical scale) multiplied by a scaling factor for graphical purposes (cf. text for details).  As explained in the text, the small shift in the peak position is attributed to environmental reasons.   Inset:  photon number PDF predicted by the Class B model (eqs.~(\ref{pofn},\ref{coeffpofn})) at the autocorrelation peak:  $g^{(2)}_0 = 2.15$ (solid line, magenta online).  For comparison, the PDF predicted shape for $g^{(2)}_0 = 1.95 < 2$ (dashed line, black online).
 }
\label{g2}
\end{figure}

Fig.~\ref{g2} shows the experimental measurement of the zero-delay second-order autocorrelation obtained from the HBT setup (blue data).  Its main feature is the appearence of a peak, at excitation current $i \approx 1.3 mA$, whose value ($g^{(2)}(0)_{peak} = 2.15 \pm 0.05$) exceeds, with three standard deviations, the thermal statistics.  The peak is very sharp and corresponds to the temporal regime of laser spikes displayed in Fig.~\ref{expthisto}a and corresponds to the bandwidth-limited peak observed with linear detection apparatus~\cite{Wang2015}.  The peak height is compatible with the bandwidth limitation known for the linear measurement and the slight difference in current value at which this feature appears is consistent with a small difference in room temperature control (the measurements displayed in Fig.~\ref{g2} and those of~\cite{Wang2015} have not been taken in the same building).  All other features are already well-known~\cite{Wang2015}, including the fact that starting from $i \geq 1.5 mA$ the laser is emitting coherent, albeit noisy, radiation~\cite{Takemura2012,Lebreton2013b}.   As usual, the thermal limit ($g^{(2)}(0) = 2$) at low pump values ($i \approx 1 \rm \, mA$) is not observed due to the temporal resolution of the coincidence measurement, but the maximum of $g^{(2)}(0)$ shows, within error bar, the emission of superthermal light~\cite{BoitierFabre2009}.

The red data (triangles) represent, for comparison with the HBT-based correlations, the $g^{(2)}(0)$ measurements obtained with the linear setup, previously published in~\cite{Wang2015}.  The limited detection bandwidth of the linear instrumentation strongly reduces the actual values of $g^{(2)}(0)$~\cite{Wang2015}, which are rescaled here for graphical purposes by plotting $1 + [g^{(2)}(0) - 1] \times 10$.  
The sharper drop in autocorrelation recostructed from the linear measurements (red triangles) is ascribed to the laser's intrinsically low dynamical stability in the corresponding current range and to the higher sensitivity of the HBT setup.

\section{Discussion}\label{discussion}

These observations carry a number of implications on our current understanding of the development of coherence in lasers.  The standard view of coherence establishment, based on class A lasers -- i.e., memoryless and (usually) macroscopic devices --, presents the transition region between fully incoherent and entirely coherent emission as a statistical superposition state where one fraction of photons is coherent, while the other is incoherent.  The weight of the coherent fraction gains in size until the full establishment of optical coherence:  this is the regime commonly accepted as {\it laser emission}.  Measurements taken in solid-state microcavities~\cite{Lien2001,Woerdman2001,vanDruten2000} refined somewhat this view, showing that the introduction of long-term memory in the system (the population variable in class B lasers) partially modify the photon statistics, without, however, changing the previous picture:  coherence is supposed to gradually emerge from the fully incoherent emission until lasing.  

Our observations show that this is not the case and hint to the existence of a new scenario, as detailed in the following sections, where the dynamical growth of coherence plays an important role.  In hindsight, the finding is not unreasonable since the intrinsic memory embedded in Class B systems is bound to emerge in some form, unlike the traditional Class A picture of the lasing transition. 

\subsection{Photon statistical implications}

Numerous physical mechanisms (quantum interference~\cite{Temnov2009,Jahnke2016}, broadband emission~\cite{Guan2012}, or multimode effects~\cite{Schlottman2017,Marconi2018}) can give rise to superthermal emission, but none of them is likely to intervene in our experiment.  Instead, the observed spontaneous pulsing dynamics~\cite{Wang2015}, which matches the pump values for superthermal emission (Fig.~\ref{expthisto}a), is the most probable origin of photon bunching.  While the phenomenology resembles that of two-mode competition~\cite{Sondermann2003,Schlottman2017,Marconi2018} which also gives rise to isolated pulses due to switching between the emission modes, the dynamics is qualitatively different as it is born from:  1. the intrinsic stochasticity in the establishment of stimulated emission; and 2. the memory effects introduced by the slow-responding gain medium.

The effect of memory alone, established for macroscopic Class B lasers~\cite{Paoli1988} and exploited in~\cite{Lien2001}, is, however, insufficient to account for the observations.
Reformulating Class B photon statistics~\cite{Paoli1988} in terms of average photon number, $n_a$, and $g^{(2)}(0)$ as in~\cite{Lien2001}, provides a correspondence between the statistical distributions (Fig.~\ref{expthisto}) and the autocorrelation measurements (Fig.~\ref{g2}):
\begin{eqnarray}
\label{pofn}
P(n) & = & C(n_a,g_{2,0}) n^{\frac{2-g_{2,0}}{g_{2,0}-1}} e^{-\frac{n}{n_a [g_{2,0} -1]}}, \\
\label{coeffpofn}
C(n_a,g_{2,0}) & = & \frac{1}{\Gamma \left( \frac{1}{g_{2,0} - 1} \right)} \left( \frac{1}{n_a (g_{2,0} -1)} \right)^{\frac{1}{g_{2,0} -1}} \, ,
\end{eqnarray}
where we have explicitely written the coefficients given in~\cite{Lien2001} and used the shorthand $g_{2,0}$ in place of $g^{(2)}(0)$; $\Gamma \left( \frac{1}{g_{2,0} - 1} \right)$ represents the Gamma-fuction of the corresponding argument.  Inspection shows immediately that the nature of the distribution changes entirely when $g^{(2)}(0) > 2$ and transforms the PDF into a monotonically decreasing distribution, as shown in the inset of Fig.~\ref{g2} (solid line), which is clearly incompatible with our experimental observations (Fig.~\ref{expthisto}a), as well as with the expected generalization of previous results~\cite{Lien2001}.   The apparent lack of bijectivity between the two sets of data highlights an underlying limitation of the model~\cite{Paoli1988} which, by  neglecting the contribution of spontaneous emission and its quantum fluctuations, becomes intrinsically unable to predict the occurrence of the dynamics observed in Fig.~\ref{expthisto}a,b.  Since the PDF collects into a statistical distribution the state of the system, its predictions fail when the dynamics cannot be reproduced by the model from which it is derived.  Different emission statistics appear therefore to produce $P(n)$ distributions whose form is indistinguishable, at least with the current measurement techniques.  The inference, accepted up until now, that the photon statistics for semiconductor--based micro- and nanolasers should remain the same as the one predicted for macroscopic lasers~\cite{vanDruten2000,Lien2001} cannot hold, otherwise we would not have met with the inconsistency in the PDF.

\subsection{Comparison to currently accepted conclusions}

This observation implies a shift of paradigm in the role that current knowledge of photon statistics plays in the physics of the laser threshold, particularly at the meso- and nanoscale.  Paraphrasing Lien {\it et al.}~\cite{Lien2001}, the {\it non-standard} Class B photon statistics, which {\it unexpectedly} matched the experimental observations, has been thought for the past two decades to hold for only lasers with $\Lambda \beta \gtrapprox 1$, following its verification on a microcavity solid state laser.  Reasonable dynamical considerations~\cite{vanDruten2000} indicated that for semiconductor devices the first deviations from standard Class A photon statistics should rather be observed for $\Lambda \beta \gtrapprox 0.1$~\cite{Lien2001}.

Thus, it was widely understood that one could rest the interpretation of statistical correlations obtained in small devices (e.g.~\cite{Strauf2006,Wiersig2009}) on standard, class A photon statistics, since, based on work completed two decades ago, the photon statistics at the laser threshold could be considered entirely known.  According to this picture, for true nanolasers (e.g.~\cite{Ota2017,Jagsch2018}), macroscopic, Class B photon statistics should apply.

Yet, our current results show that already for $\Lambda \beta \lessapprox 10^{-2}$ Class B photon statistics~\cite{Paoli1988} are in fair agreement with the experimental observations.  This highlights the role played by memory effects even when the characteristic timescales of photons and reservoir differ
only by a couple of orders of magnitude (in solid state microdisk devices the ratio is four orders of magnitude or more~\cite{vanDruten2000,Woerdman2001,Lien2001}).  In addition, the observation of superthermal emission introduces an entirely new statistical regime in need of physical explanation. 

Deviations from {\it standard} photon statistics were expected close to threshold at the small scale, and were the motor behind the investigations carried out on microscale solid-state lasers, as surrogate for semiconductor-based nanolasers.  However, the experimental observations~\cite{Lien2001} led towards the already predicted distributions~\cite{Paoli1988} and to the conclusion that all nanodevices would follow these same statistics.  

Thus the question naturally arises as to why the dynamics of Fig.~\ref{expthisto}a were not previously observed~\cite{Woerdman2001}.  The most likely reason is that in spite of the relatively large $\beta$ achieved in a solid state microcavity ($7 \times 10^{-6} \lessapprox \beta \lessapprox 2 \times 10^{-5}$~\cite{vanDruten2000}), its volume remained too large to provide a sufficiently broad parameter region (pump rate range) to experimentally access the pulsing regime.  In fact, as the steepness of the laser S-shaped response curve sensitively depends on  $\beta^{-1}$, the range of pump rate values in which such dynamics can be observed shrinks, as confirmed by stochastic numerical simulations based on~\cite{Puccioni2015}, rendering its experimental observation quite difficult.  In addition, the {\it extreme Class B} nature of the solid state microcavity renders the observation of such dynamics even harder since the strong imbalance between cavity losses and population inversion relaxation rate makes the device more sensitive to pump fluctuations~\cite{Lien2001}.  This hypothesis is further strenghtened by the fact that the regime of strong oscillations (cf. Fig.~\ref{expthisto}c) was indeed reported~\cite{Woerdman2001}, but no spiking dynamics appears to have been observed, presumably due to the narrowness of the pump interval over which the phenomenon would appear.  This explains the lack of this crucial observation which lays at the root of the disagreement between previous conclusions~\cite{Lien2001,vanDruten2000,Woerdman2001} and our current observations.

\subsection{Progress through dynamical description}\label{dynprog}

One important aspect of revisiting the threshold photon statistics of small lasers resides in its intrinsic usefulness when characterizing nanolasers.  The very low flux associated with nanodevices renders the autocorrelation function (in particular $g^{(2)}(0)$~\cite{Kreinberg2017}) the ideal discerning tool, thanks to the extremely high sensitivity of single photon counting detectors and the large bandwidth of current associated electronics.  However, the interpretation of the resulting statistics requires a good knowledge of the dynamical behaviour which underpins the observations, since similar statistical distributions may be obtained from signals with intrinsically different features.  The advantage offered by our choice of a mesoscale device, with its moderate $\beta \approx 10^{-4}$ value, is that it is possible to obtain temporal information, reconstruct statistical distributions (albeit filtered, as discussed in section~\ref{linear}) and measure the autocorrelation (directly, through quantum coincidences, and indirectly, from the time series).

\begin{figure}[htbp]
\centering
\includegraphics[width=\linewidth,clip=true]{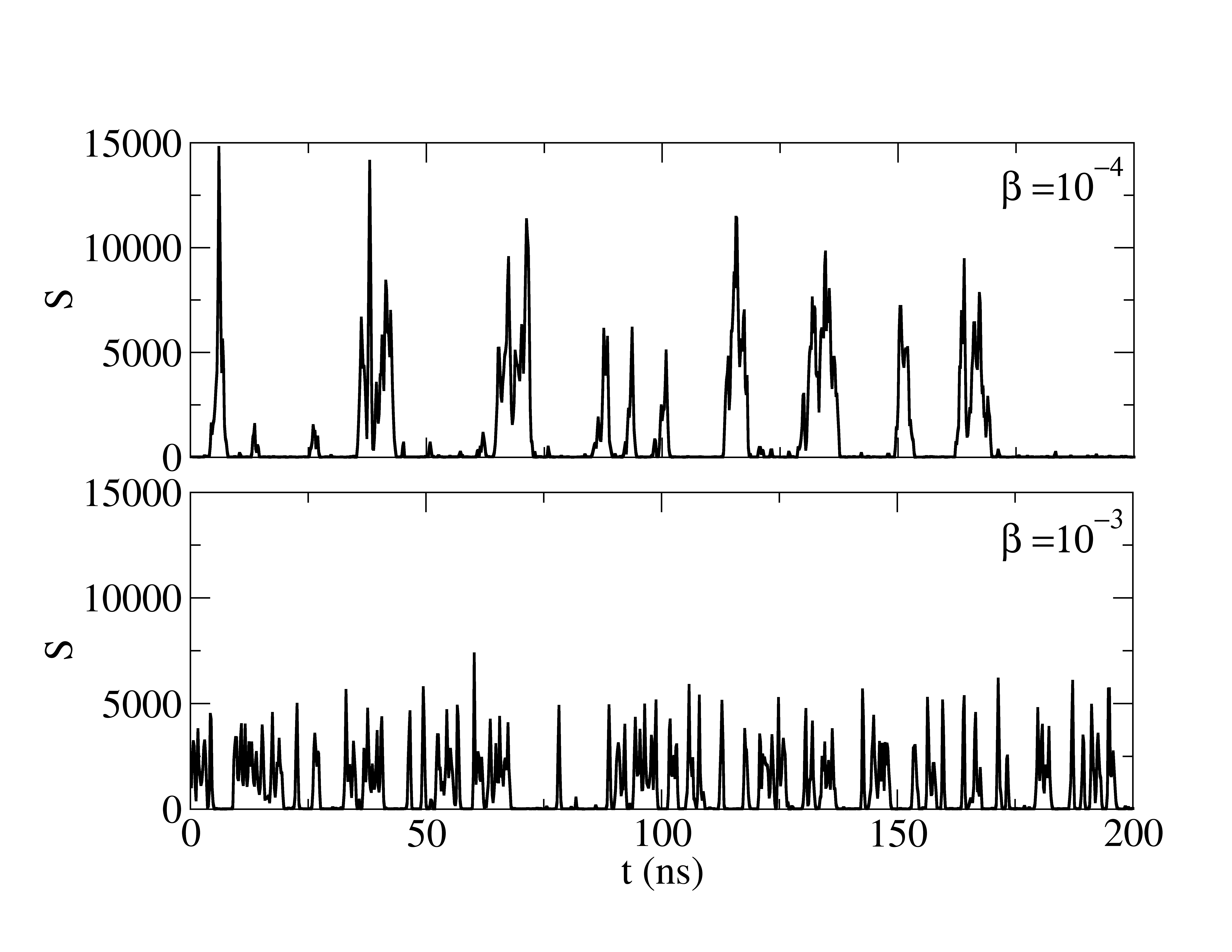}
\caption{
Photon number computed from the Stochastic Simulator for a laser with $\beta = 10^{-3}$ (bottom panel) and $\beta = 10^{-4}$ (top panel).  A detector integration time $\Delta t_{det} = 0.1 ns$ has been included to average the fast fluctuations coming from the high-resolution simulations (time step $t_s = 5 \times 10^{-14} s$).  $P = 0.99 \times P_{th}$ for both panels.  Details on the computations can be found in~\cite{Puccioni2015}, where all other parameter values are given.
}
\label{autocorr-beta}
\end{figure}

Additional understanding of the current observations can be gained with the help of the fully stochastic laser simulation~\cite{Puccioni2015}.  
The laser dynamics in the threshold region (at the peak of the autocorrelation, Fig.~\ref{g2}) computed by the laser simulator is displayed in Fig.~\ref{autocorr-beta} (top panel).  It consists of sharp bursts of emission with peak values about two orders of magnitude larger than their average.  The origin of these photon bursts lies in the intrinsic memory introduced by the slow energy reservoir:  the carriers evolve slowly, compared to the photon number, and can accumulate an excess number of carriers due to the stochastic nature of the conversion into stimulated emission.  The process is similar to that observed in gain-switched devices where a large spike of emission takes place following the accumulation of excess energy in the reservoir.  However, since the pump is insufficient to ensure continuous emission, the photon number drops to zero, letting the carriers accumulate again.  As visible in the figure, photon bursts are not the only possible kind of emission, at times small numbers of photons can also exit the cavity (very low peaks visible, for instance $t = 125 ns$).
 
This dynamics, which cannot be reproduced by standard rate-equation models even with the inclusion of Langevin noise, is responsible for the observed {\it photon bunching} which produces the superthermal statistics.  In addition, we see that contrary to~\cite{Schlottman2017,Marconi2018} there is no need for the competition with a second mode, since the photon bursts originate from the memory effects which are intrinsic to the non-markovian nature of Class B devices.  Of course, the absolute value of $g^{(2)}(0)$ is not as high as in experiments with mode competition, where the superthermal nature of the statistics for the suppressed mode originates from the rare switching events.  Instead, the single-mode threshold behaviour we are observing derives its dynamics from the intrinsic physical interactions between photons and energy reservoir in a way which has so far remained unexplored at threshold and which is not included in current analytical models.

\subsection{Scaling of autocorrelation maximum with $\beta$ parameter}

Autocorrelation measurements conducted on a smaller micro-VCSEL (Ulm Photonics, Single Mode VCSEL 980 nm) with an estimated $\beta \approx 10^{-3}$, with the apparatus used for the quantum coincidence measurements (Section~\ref{quantummeas}) give a functional dependence of $g^{(2)}(0)$ very similar to the one observed in Fig.~\ref{g2}.  The main difference is the height of the maximum which, instead of giving a superthermal emission, is limited to $g^{(2)}_{max}(0) \approx 1.6$.  The statistical result alone begs the question of whether spiking exists only in a very restricted interval of $\beta$-parameter values.  Indeed, one could interpret the subthermal value of $g^{(2)}_{max}(0)$ as the convolution between the instrumental response, which filters the autocorrelation to the shot-noise value ($g^{(2)}(0) = 1$) at very low pump due to limited resolution, and the physically relevant progressive coherence growth (down from 2 towards the Poisson limit).  

If this were the case, then the progression from thermal to Poisson emission in smaller mesoscale and nanolasers may proceed as {\it normally}  expected, through a progressive increase of the statistical weight of the stimulated photon fraction.  This question holds weight in the interpretation, and in the physical description, of measurements conducted in nanolasers when {\it imperfect coherence} is measured.

For the device in question, in spite of the low-resolution in photon number due to the lower flux coming from the smaller device, we still have access to time-resolved measurements dynamical measurements which show traces of self-spiking, as the one displayed in Fig.~\ref{expthisto}.  Even more important, however, is the contribution of the stochastic simulations, which can be extrapolated to the nanoscale.

Fig.~\ref{autocorr-beta} compares a predicted temporal emission sequence for a laser with $\beta$-value (bottom panel) compatible with the smaller micro-VCSEL analysed in this subsection to one for a larger device (top panel), for which the superthermal emission clearly indicates an {\it anomalous} statistical behaviour (Fig.~\ref{g2}).  The dynamics is in both cases composed of photon bursts, but for the smaller laser the spikes are typically closer to one another with fewer {\it interruptions} in the emission; at the same time, the amplitude is smaller.  In any case, photon bursts are emitted by the smaller laser, as well as by nanolasers (cf.~\cite{Puccioni2015} and Supplementary Information in~\cite{Wang2015}).  Thus, the difference is quantitative, rather than qualitative.  The temporal sequence (bottom panel of Fig.~\ref{autocorr-beta}) also explains the reason why $g^{(2)}_{max}(0)$ would be smaller for the larger $\beta$ laser:  the lower peaks and the (relatively) larger average coming from the more frequent spikes reduces the value of the autocorrelation maximum, which from the numerical simulations give 
\begin{eqnarray}
\left. g^{(2)}_{\beta = 10^{-3}}(0) \right|_{num}  \approx  0.763  \left. g^{(2)}_{\beta = 10^{-4}}(0) \right|_{num} \approx 1.63\, ,
\end{eqnarray}
in good agreement with the experimental observations.

Thus, the message which comes from the scaling properties is that the photon dynamics which precedes cw emission consists of photon bursts, independently of cavity size, but that the autocorrelation measurement alone gives results which do not clearly permit its identification (since a subthermal peak could be interepreted as a convolution between instrumental and physical effects).  Only the dynamical measurements, compared to stochastic simulations, can clearly identify the presence of self-spiking.

The relevance of these considerations becomes immediate when compared to published experimental observations which generically show the same features we are reporting here.  Peaks in $g^{(2)}(0)$ are reported in a $\beta \approx 0.7$ nanobeam laser~\cite{Jagsch2018} with $g^{(2)}(0) < 1.5$ at its peak.  While no direct measurement of the temporal evolution of the photon number is available for that experiment, it is possible that the peak may be the consequence of similar dynamics (numerically already predicted, cf. Supplementary Information in~\cite{Wang2015}).

Summarizing, we find that the stochastic dynamics provides precious and for the moment irreplaceable information on the origin of the autocorrelation value to identify its relationship to actual field coherence.  The ability to directly measure the temporal evolution of the emitted laser field intensity and to correlate it to statistical measurements represents a sizeable step forward in our understanding of new features which appear in the establishment of coherent emission in small (and not so small) lasers, modifying the currently accepted picture of a statistical, time-independent superposition of coherent and incoherent photons which evolves by changing the relative weight of the two components as the energy provided to the laser brings the latter above its {\it emission} (or coherent) {\it threshold}.

\subsection{Pump range scaling for photon bursts}

The previous considerations may indicate that photon bursts should be easier to detect in macroscopic lasers than in their smaller counterpart, thus questioning the reason for their lack of observation in carefully conducted experiments~\cite{vanDruten2000,Lien2001,Woerdman2001}.  The question is best addressed by turning again towards stochastic simulations which allow for an exploration of the pump range in which bursts appear.  The definition of the pump interval in which self-spiking appears requires some degree of interpretation; the intrinsic stochasticity of the phenomenon renders the identification of the boundary uncertain.  Since we are interested in finding a qualitative scaling, a fine scan of the pump provides sufficient information.

\begin{figure}[htbp]
\centering
\includegraphics[width=\linewidth,clip=true]{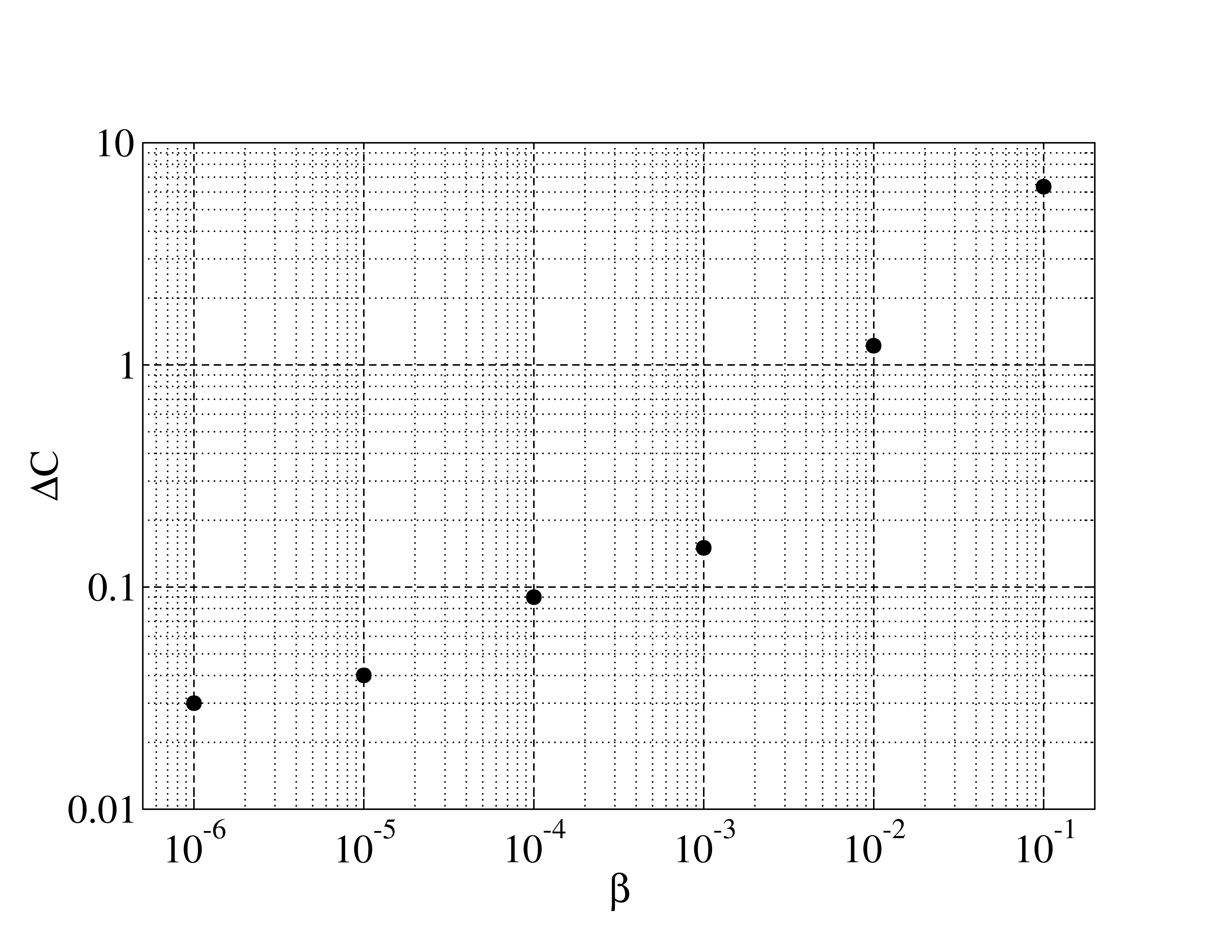}
\caption{
Width of the pump interval, normalized to threshold, over which a spiking output is observed as a function of $\beta$, computed with stochastic simulations~\cite{Puccioni2015}.  Passing from $\beta = 10^{-4}$ to $\beta = 10^{-6}$ we find a reduction in the pump interval of approximately a factor 3, rendering the observation of self-spiking much more difficult.
}
\label{pumpint}
\end{figure}

Fig.~\ref{pumpint} displays, in double logarithmic scale, the width of the pump interval (normalized to its threshold value) $\Delta C$ over which photon bursts are observed as a function of cavity $\beta$.  The functional behaviour -- whose irregularity is to be attributed to the above-mentioned difficulty in identifying the true pump boundaries -- shows a clear growth with $\beta$ with a reduction in width by several units (possibly approaching one order of magnitude) when passing from $\beta = 10^{-4}$ to $\beta = 10^{-6}$.  Comparison to Fig.~\ref{expthisto}, which shows an already narrow interval ($\Delta i < 0.1$ mA) for $\beta = 10^{-4}$, lends support to the conjecture that even very careful experiments conducted with extremely good parameter stabilization would fail to reliably detect photon bursts.  Even in our larger micro-VCSEL great care has been exercised with the use of very stable current supplies, of high-quality temperature stabilization and by isolating the device from air currents (double box on a floating optical table).  The integrated nature of the semiconducting laser naturally contributes to an efficient isolation from external perturbations.

In addition, the large difference in relaxation rates typical of solid-state devices, which enable values for $\Lambda$ of the order of 100~\cite{Lien2001}, play against intrinsic stability.  The long time constants, beneficial in the detection bandwidth and sensitivity, render the device more sensitive to slow thermal fluctuations.  In addition, the large ratio between the photon decay rate and the population (up to 6 or 7 orders of magnitude in~\cite{Lien2001}) enables very small cavity fluctuations to substantially perturb the whole laser dynamics.  It is therefore not surprising that even an excellently designed and run experiment may have failed to detect the presence of (unexpected) photon bursts.

Finally, it is worth remarking that if the stochastic predictions hold, then in a $\beta = 0.1$ nanolaser the self-spiking interval should reach $P \approx 7 \times P_{th}$, with substantial consequences on the functional dependence of $g^{(2)}(0)$ in such devices and on the interpretation of the statistical significance of coherence in these nanolasers.

\subsection{Issues related to photon statistics and spiking in different dynamical classes}

We have connected the discrepancies between the photon statistics observed in a Class B micro-VCSEL and the predictions stemming from models to the existence of a self-spiking regime before the emission becomes cw (albeit very noisy).  As already mentioned, this regime is responsible for the sharp peak in the second-order autocorrelation function but is not the sole cause of discrepancy.  The intrinsic memory present in a Class B laser, stemming from the slow population response, adds to the complexity of the problem.

The existence of spiking has been predicted by statistical models close to Class A devices~\cite{Roy2009,Roy2010} and has been confirmed by more sophisticated Monte-Carlo simulation taking into account the electronic band dynamics of a semiconductor~\cite{Vallet2019} (in Class A regime up to the Class B limit, represented by equal relaxation constants for carriers and photons).  Their origin bears a resemblance to the one which explains their appearence in Class B lasers, but for the exchange in the role of slow and fast time constants~\cite{Wang2019b}.  

The validity of standard photon statistics is predicted for Class A devices (up to the marginal regime between Class A and Class B examined in~\cite{Vallet2019}) by analyzing higher order correlations $g^{n}(0)$ with $n \le 10$~\cite{Takemura2019b}. The same analysis, however, confirms the anomaly of the true Class B regime, as analyzed in this paper, whose statistics does not conform to the Class A one.  Specifically, the case of $\Lambda = 10^{-2}$ has been considered there and has shown to possess visible discrepancies, thus confirming the lack of validity of the extrapolations to Class-B nanolasers based on solid-state microresonator results~\cite{Lien2001}.

\subsection{Outlook}

The discussion of the experimental observations presented in this paper clearly shows that current analytical models are incapable of capturing some of the features of the laser behaviour in its threshold region, and therefore of providing dependable photon statistical models on which to base the interpretation of measurements conducted in nanolasers.  
The challenges posed by a more complete modeling approach are considerable, since the latter must cover the intermediate ground where neither the fully quantum description is applicable, nor the thermodynamic limit can be taken.  Full quantum calculations have been already performed (e.g., \cite{Temnov2009,Chow2014,Jahnke2016,Gies2007}), but are limited to a few electromagnetic cavity modes ($\beta > 0.1$) and few emitters.  On the other hand, the averaging process which allows for the macroscopic description (thermodynamic limit) appears to hold sufficiently well only for $\beta < 10^{-4}$ at least in the transition region from incoherent to coherent emission.  It is in this gap, $10^{-4} \le \beta \le 0.1$, that the new model needs to hold, taking into account in some form the stochastic nature of the quantum-mechanical spontaneous emission, and the emergence of spontaneous spikes~\cite{Wang2015,Wang2017,Puccioni2015,Vallet2019}.  While interesting {\it ad hoc} procedures are being applied to explain the photon statistics with mode competition~\cite{Lettau2018}, for the more general case novel approaches, such as~\cite{Redlich2016}, will most likely be needed to overcome the limitations of the usual approximations and contribute to a better understanding of the physics and operation of nanolasers.
Indeed, since next-generation photonics integrated circuits can be expected to be based on nanolasers operating in the few-photon regime, a good understanding of their photon statistics might prove a significant technological issue in the near future.

\section{Conclusions }
In summary, this paper shows three new results and conclusions:  the photon statistics of semiconductor micro- (and, by extension, nano-cavities) does not follow class A predictions, contrary to previous proof~\cite{Lien2001}; superthermal light is spontaneously emitted as a consequence of intrinsic dynamics in single-mode microcavities with $\beta \approx 10^{-4}$;  new and more sophisticated models are required to predict the photon statistics of small-scale lasers ($10^{-4} \lessapprox \beta \lessapprox 0.1$) in the crucial region at the transition between incoherent and coherent emission.  We have further shown that knowledge of the dynamics plays a crucial role in the identification of the laser features, since same statistical features may correspond to entirely different emission characteristics.

We stress that these results hold a potentially strong impact for nanolasers, where at the present time only photon statistical measurements can be carried out.  Recent experimental observations have shown that in a high-$\beta$ laser ($\beta = 0.22$, cf. Fig. 4, blue line, in~\cite{Ota2017}) the zero-delay second-order autocorrelation possesses a plateau similar to  the one repeatedly found in mesoscale VCSELs, thus confirming the existence of common features between devices at the meso- and at the nanoscale.

\begin{acknowledgments}
Partial funding was obtained from the F\'ed\'eration W. D\"oblin.  GLL acknowledges enlightening discussions with A. Politi.  T.W. is grateful to the Conseil R\'egional PACA for funding and to BBright for support.  The authors acknowledge support from the French government through its Investments for the Future programme under the Universit\'e C\^ote d'Azur UCA-JEDI project (Quantum@UCA) managed by the ANR (ANR-15-IDEX-01).
\end{acknowledgments}

\end{document}